\documentclass[prb,amsmath,amsfonts,amssymb,twocolumn]{revtex4}
\usepackage{graphicx}
\usepackage{bm}
\usepackage{color}
\begin{document}
\bibliographystyle{apsrev}
\title{ROHF Theory Made Simple}
\author{Takashi Tsuchimochi$^*$ and Gustavo E. Scuseria$^{* \dagger}$}
\affiliation{$^*$Department of Chemistry, Rice University, Houston, TX 77005-1892\\
			$^\dagger$Department of Physics and Astronomy, Rice University, Houston, TX 77005-1892}

\begin{abstract}
Restricted open-shell Hartree-Fock (ROHF) theory is formulated as a projected self-consistent unrestricted HF (UHF) model by mathematically constraining spin density eigenvalues. The resulting constrained UHF (CUHF) wave function is identical to that obtained from Roothaan's effective Fock operator. Our $\alpha$ and $\beta$ CUHF Fock operators are parameter-free and have canonical orbitals and orbital energies that are physically meaningful as in UHF, except for eliminating spin contamination. The present approach removes ambiguities in ROHF orbital energies and the non-uniqueness of methods that build upon them. We present benchmarks to demonstrate CUHF physical correctness and good agreement with experimental results.
\end{abstract}
\maketitle

Restricted open-shell Hartree-Fock (ROHF) theory was formulated by Roothaan some 50 years ago.\cite{Roothaan} A major drawback of this model is the lack of a unique effective Fock operator.\cite{PGB} Even though the ROHF wave function and total energy obtained from different coupling schemes are the same, the resulting orbitals and orbital energies are different and lead to post-ROHF results that generally depend on them. The interpretation and physical picture emerging from Roothaan's open-shell theory have always been somewhat blurry. Attempts to resolve these ambiguities, as well as many paradoxes resulting from them, are well documented in the literature.\cite{PD-JMC09,PD-JPCA09,DP-JCP09,GS10}

On the other hand, the physical picture of unrestricted HF (UHF) is clear.\cite{UHF} It is a single-determinant wave function with well-defined $\alpha$ and $\beta$ orbital energies obeying Koopmans' theorem. It is straightforward to use it in post-UHF calculations by simply treating the $\alpha$ and $\beta$ orbitals explicitly and separately. The notorious problem in UHF, however, is spin contamination: the wave function is not an eigenfunction of $S^2$. This weakness is ubiquitous and a serious detriment when bonds are stretched. If the UHF wave function suffers from severe spin contamination, as is the case when strong static correlation is present, then UHF is no longer a good starting reference point for post-UHF treatments of correlation or excited states. Once lost, good quantum numbers are hard to recover,\cite{BHS} so when possible, it is preferrable to use ROHF as a starting point despite the ambiguities regarding its associated Fock operator.

In recent work, we have developed a novel theory for treating strong correlations within an independent quasiparticle picture.\cite {CPMFT1,CPMFT2,CPMFT3,CPMFT4} As a spin off of this work,\cite{CPMFT4} we have realized that the UHF energy can be written as a functional of the charge density matrix $\mathbf{P}=(\bm{\gamma}^\alpha+\bm{\gamma}^\beta)/2$ and the spin density matrix $\mathbf{M}=(\bm{\gamma}^\alpha-\bm{\gamma}^\beta)/2$, where $\bm{\gamma}^\alpha$ and $\bm{\gamma}^\beta$ are the $\alpha$ and $\beta$ density matrices, respectively. Our proposed Constrained-Pairing Mean-Field Theory (CPMFT) departs from UHF by adopting a different definition for $\mathbf{M}$, a choice inspired by an underlying quasiparticle correlation picture. With this choice, CPMFT can accurately and efficiently describe static correlation and dissociate any molecule to fragments with ROHF energies, keeping the correct $\langle S^2 \rangle$ all along the dissociation path.\cite{CPMFT2} The connections between CPMFT, ROHF, and UHF turn out to be enlightening for formulating ROHF as a constrained UHF theory. The resulting CUHF scheme here presented leads to well-defined $\alpha$ and $\beta$ Fock operators with straightforward interpretation and no spin contamination. The ROHF wavefunction, energy, charge and spin densities remain the same; only the ROHF Fock operator is replaced by two UHF-like counterparts. As shown in benchmarks below, the meaning of the resulting orbitals and orbital energies is much more physical than in Roothaan's approach and provide a base for treatments of electron correlation and excited states.
 
{\it Theory.} The familiar energy expression in ROHF is
\begin{equation}
E_{\rm ROHF}=2\sum_i f_i h_{ii} + \sum_{ij}f_i f_j (2a_i^j\langle ij|ij\rangle - b_i^j\langle ij|ji \rangle),
\end{equation}
where $h_{ij}$ are one-electron integrals,
$\langle ij | kl \rangle$ are two-electron integrals in Dirac's notation, $a$ and $b$ are the coupling coefficients, and $f_i$ are the orbital occupations: 1 for core (doubly-occupied, $c$) and 0 for virtual (unoccupied, $v$) orbitals. In the case of high-spin open-shell systems under consideration, $a=1$, $b=2$, and $f=1/2$ for open-shells orbitals (singly-occupied, $o$). Roothaan's effective Fock operator is defined as
\begin{eqnarray}
\mathbf{F}_{\rm ROHF}=\left(
\begin{array}{lll}
\mathbf{R}_{cc}           &   \mathbf{F}^\beta_{co}    & \mathbf{F}^{\rm cs}_{cv} \\
\mathbf{F}^\beta_{oc}     &   \mathbf{R}_{oo}          & \mathbf{F}^\alpha_{ov} \\
\mathbf{F}^{\rm cs}_{vc}  &   \mathbf{F}^\alpha_{vo}   & \mathbf{R}_{vv} 
\end{array}\right)
\begin{array}{l}
\mbox{core ($c$)}\\ \mbox{open ($o$)}\\ \mbox{virtual ($v$)}
\end{array}
\label{eqn:Roothaan}
\end{eqnarray}
where $\mathbf{F}^\alpha $ and $\mathbf{F}^\beta$ are UHF $\alpha$ and $\beta$ Fock matrices, and $\mathbf{F}^{\rm cs} = (\mathbf{F}^\alpha+\mathbf{F}^\beta)/2$. At self-consistent field (SCF) convergence, all off-diagonal $\mathbf{F}_{\rm ROHF}$ terms become zero. The choice of the diagonal elements in Eq.(\ref{eqn:Roothaan}) is completely arbitrary within a set of $A$ and $B$ coupling parameters:
\begin{subequations}
\begin{align}
\mathbf{R}_{cc}&=A_{cc}\mathbf{F}^\alpha_{cc}+B_{cc}\mathbf{F}^\beta_{cc}\\
\mathbf{R}_{oo}&=A_{oo}\mathbf{F}^\alpha_{oo}+B_{oo}\mathbf{F}^\beta_{oo}\\
\mathbf{R}_{vv}&=A_{vv}\mathbf{F}^\alpha_{vv}+B_{vv}\mathbf{F}^\beta_{vv}.
\end{align}
\end{subequations}
Different values for these parameters have been suggested in the literature.\cite{PGB} Although they do not affect the ROHF wave function and energy, they affect orbital energies whose physical meaning is obscured because of this dependence. Choices guided to determine ``canonical" sets that satisfy Koopmans' theorem may result in violations to the aufbau principle.\cite{PD-JPCA09,GS10} In this paper, these problems are resolved by abandoning the use of a single Fock operator. We will obtain the ROHF wave function by \textit{projecting} the UHF wave function self-consistently. Spin contamination in UHF is given by\cite{CPMFT4}
\begin{equation}
\delta_s = \langle S^2\rangle - S_z(S_z+1) = N_\beta - \mathrm{Tr}(\bm{\gamma}^\alpha\bm{\gamma}^\beta)
\end{equation}
where  $S_z=(N_\alpha-N_\beta)/2$ and  $N_\sigma$ ($\sigma = \alpha , \beta$) is the number of $\sigma$ electrons in the system. The previously proposed spin-constrained UHF (SUHF) approach\cite{SUHF} introduces a Lagrange multiplier $\lambda$ in UHF to enforce $\delta_s=0$. However, this is exact only in the limit of $\lambda\rightarrow \infty$. In this limit, the effective SUHF Fock matrices remain in the form of Eq.(\ref{eqn:Roothaan}).\cite{ROMP2} We here propose an alternative method based on restricting natural occupations and spin density eigenvalues via finite Lagrange multipliers. 

In UHF, the natural occupations $n$ are eigenvalues of $\mathbf{P}$; they can be 0, 1, $\frac{1}{2}$, or appear in ``corresponding pairs" $(n,1-n)$.\cite{Harriman} This is a rigorous mathematical result following from $\mathbf{P}$ being the half sum of two idempotent density matrices.\cite{math} In high-spin systems, the number of $\frac{1}{2}$ occupations is  $N_\alpha-N_\beta=N_s$ (we assume $N_\alpha>N_\beta$ always). Note that $\mathrm{Tr}\mathbf{P}= (N_\alpha + N_{\beta})/2 = N_e/2$, where $N_e$ is the number of electrons. For clarity, we discuss below only the case where the number of orbitals $N$ is greater than $N_e$ but our results hold for $N \le N_e$ too. The UHF $\bm{\gamma}^\sigma$ are block-diagonal in the NO basis:
\begin{equation}
\bm{\gamma}^\alpha = 
\begin{pmatrix} 
 \bm{\gamma}_1^\alpha& & & &\\
  & \ddots & & &\\ 
&&\bm{\gamma}_{N_{cp}}^\alpha&&\\
&&&\bm{1}&\\
&&&&\bm{0}\\
\end{pmatrix},
\hspace{0.2cm}
\bm{\gamma}^\beta = 
\begin{pmatrix} 
 \bm{\gamma}_1^\beta& & & &\\
  & \ddots & & & \\ 
&&\bm{\gamma}_{N_{cp}}^\beta&&\\
&&&\bm{0}&\\
&&&&\bm{0}\\
\end{pmatrix}
\label{eqn:PinNO}
\end{equation}
where $N_{cp}$ is the number of corresponding pairs and
\begin{equation}
\bm{\gamma}_i^{\alpha}=
\begin{pmatrix} n_i & +m_i \\   +m_i & 1-n_i \end{pmatrix},\:\:
\bm{\gamma}_i^{\beta}=
\begin{pmatrix} n_i & -m_i \\   -m_i & 1-n_i \end{pmatrix}\label{eqn:Pi}
\end{equation}
and $m_i = \sqrt{n_i-n_i^2}$. 
The identity matrix in $\bm{\gamma}^\alpha$ accounts for unpaired electrons, traces to $N_s$, and is substituted by a corresponding zero matrix in $\bm{\gamma}^\beta$. The other zero matrix has dimension $N_v=N-N_s-2N_{cp}$ and corresponds to virtual ($n=0$) unpaired orbitals. 
In the NO basis, $\mathbf{M}$ is
\begin{equation}
\mathbf{M} = 
\begin{pmatrix} 
 \mathbf{M}_1& & & &\\
  & \ddots & & &\\ 
&&\mathbf{M}_{N_{cp}}&&\\
&&&\frac{1}{2}\cdot\bm{1}&\\
&&&&\bm{0}\\
\end{pmatrix},
\label{eqn:M}
\end{equation}
where, from Eq.(\ref{eqn:Pi}), $\mathbf{M}_i= (\bm{\gamma}_i^\alpha-\bm{\gamma}_i^\beta)/2$ is
\begin{equation}
\mathbf {M}_i=
\begin{pmatrix} 0 & m_i \\  m_i & 0 \end{pmatrix}\label{eqn:Mi}
\end{equation}
which is traceless with eigenvalues $\pm m_i$. The full spectrum of $\mathbf{M}$ also includes $\frac{1}{2}$ and 0 eigenvalues, thus tracing to $N_s/2$. Using the idempotency of $\bm{\gamma}^\alpha$ and $\bm{\gamma}^\beta$, we get 
\begin{equation}
\mathrm{Tr}(\bm{\gamma}^\alpha\bm{\gamma}^\beta)=\frac{N_e}{2}-2\;\mathrm{Tr}\mathbf{M}^2.
\end{equation}
Considering Eqs.(\ref{eqn:M}) and (\ref{eqn:Mi}), it is evident that
\begin{equation}
\mathrm{Tr}\mathbf{M}^2 = 2\sum_i^{N_{cp}}m_i^2+\frac{N_s}{4},
\end{equation}
and hence
\begin{equation}
\delta_s = N_\beta-\mathrm{Tr}(\bm{\gamma}^\alpha\bm{\gamma}^\beta)= 4\sum_i^{N_{cp}}m_i^2.
\label{eqn:Nb-TrPaPb1}
\end{equation}
This readily means that to eliminate spin contamination in UHF all $m_i$ should be zero. Therefore, we propose to formulate ROHF as a constrained UHF scheme that enforces all $m_i$ to be zero. From Eq.(\ref{eqn:Pi}), $m_i=0$ implies that corresponding pair occupations become constrained to values of 1 and 0, thus effectively creating core ($c$) and virtual ($v$) orbital blocks. To enforce these constraints, we introduce Lagrange multipliers $\lambda_{ij}$ and then write in a general basis
\begin{equation}
E_{\rm CUHF} = E_{\rm UHF} + {\sum_{ij}} '\lambda_{ij}M_{ij},\label{eqn:EROHF}
\end{equation}
where the prime on the summation restricts it to $cv$ and $vc$ blocks. $\mathbf{M}$ is unconstrained in the $oo$ block and zero in other blocks. We next derive equations for $\lambda_{ij}$.

The UHF energy is normally written as a functional of $\bm{\gamma}^{\alpha}$ and $\bm{\gamma}^{\beta}$. In our recent paper,\cite{CPMFT4} we have shown that the UHF energy expression can be alternatively written as a functional of $\mathbf{P}$ and $\mathbf{M}$,
\begin{subequations}
\begin{align}
&E_\mathrm{UHF} = E_\mathrm{cs} + E_c,
\\
&E_\mathrm{cs}  = 2 \sum_{ij} h_{ij} P_{ij} + \sum_{ijkl} (2 \langle ij | kl \rangle - \langle ij | lk \rangle )P_{ik} P_{jl}
\\
&E_c            = -\sum_{ijkl}\langle ij | lk \rangle M_{ik} M_{jl}.
\end{align}
\label{Eqn:EUHF}
\end{subequations}
$E_\mathrm{cs}$ is the closed-shell energy expression given in terms of $\mathbf{P}$, while $E_c$ is a ``correlation energy" given in terms of $\mathbf{M}$. The derivatives of $E_{\rm cs}$ with respect to $\bm{\gamma}^\alpha$ and $\bm{\gamma}^\beta$ yield the usual closed-shell Fock matrix 
\begin{equation}
            \frac{\partial E_\mathrm{cs}}{\partial \gamma^\alpha_{ij}\hfill} = 
            \frac{\partial E_\mathrm{cs}}{\partial \gamma^\beta_{ij}\hfill} = 
\frac{1}{2} \frac{\partial E_\mathrm{cs}}{\partial P_{ij}\hfill} = 
F_{ij}^\mathrm{cs}.
\end{equation}
On the other hand, the derivatives of $E_{\rm c}$ are
\begin{equation}
            -\frac{\partial E_\mathrm{c}}{\partial \gamma^\alpha_{ij}\hfill} = 
            \frac{\partial E_\mathrm{c}}{\partial \gamma^\beta_{ij}\hfill} = 
\sum_{kl}\langle ik | lj\rangle M_{kl}
\equiv \Delta^{\rm UHF}_{ij}
.
\end{equation}
Hence, 
\begin{subequations}
\begin{align}
\mathbf{F}^\alpha&=\mathbf{F}^{\rm cs}-\bm{\Delta}^{\rm UHF}\\
\mathbf{F}^\beta&=\mathbf{F}^{\rm cs}+\bm{\Delta}^{\rm UHF},\label{eqn:FUHF}
\end{align}
\end{subequations}
which are the usual UHF Fock matrices. Now, the CUHF Fock matrices additionally require the derivatives of the constraints in Eq.(\ref{eqn:EROHF}) with respect to $\mathbf{\gamma}^\alpha$ and $\mathbf{\gamma}^\beta$, which are trivially $\lambda_{ij}/2$ and $-\lambda_{ij}/2$, respectively. Defining $\bm{\Delta}^{\rm CUHF}$ as
\begin{equation}
\Delta^{\rm CUHF}_{ij} \equiv \left\{\begin{array}{cl}
\Delta^{\rm UHF}_{ij}-\frac{\lambda_{ij}}{2}   &  \mbox{ if }\{i\in c \wedge j\in v\},\\ 
& \mbox{ or }\{i\in v \wedge j\in c\}\\
\\
\Delta^{\rm UHF}_{ij}   &   \mbox{ otherwise}
\end{array}\right.
\label{eqn:DROHF}
\end{equation}
yields the CUHF $\alpha$ and $\beta$ Fock matrices,
\begin{subequations}
\begin{align}
\tilde{\mathbf{F}}^\alpha&=\mathbf{F}^{\rm cs}-\bm{\Delta}^{\rm CUHF}\\
\tilde{\mathbf{F}}^\beta&=\mathbf{F}^{\rm cs}+\bm{\Delta}^{\rm CUHF}
\end{align}
\label{eqn:FROHF}
\end{subequations}
The CUHF equations to solve are $[\tilde{\mathbf{F}}^\alpha,\bm{\gamma}^\alpha]=0$ and $[\tilde{\mathbf{F}}^\beta,\bm{\gamma}^\beta]=0$. 
Subtracting these two SCF conditions and dividing it by 2 yields
\begin{equation}
\mathbf{F}^{\rm cs}\mathbf{M}-\mathbf{M}\mathbf{F}^{\rm cs}-\bm{\Delta}^{\rm CUHF}\mathbf{P}+\mathbf{P}\bm{\Delta}^{\rm CUHF}=0.
\label{eqn:SCFcond}
\end{equation}
Partitioning these matrices into core, open, and virtual blocks gives,
\begin{subequations}
\begin{align}
&\mathbf{F}^{\rm cs}_{co}+\bm{\Delta}^{\rm CUHF}_{co}=\tilde{\mathbf{F}}^{\beta}_{co}=0\label{eqn:SCFcond2a}\\
&\mathbf{F}^{\rm cs}_{vo}-\bm{\Delta}^{\rm CUHF}_{vo}=\tilde{\mathbf{F}}^{\alpha}_{vo}=0\label{eqn:SCFcond2b}\\
&\bm{\Delta}^{\rm CUHF}_{cv}=0,\label{eqn:SCFcond2c}
\end{align}
\label{eqn:SCFcond2}
\end{subequations}
where we have used $\mathbf{P}_{cc}=\mathbf{1}$, $\mathbf{P}_{vv}=\mathbf{P}_{cv}=\mathbf{P}_{co}=\mathbf{P}_{vo}=0$, and $\mathbf{P}_{oo}= \mathbf{M}_{oo}= \frac{1}{2}\mathbf{1}$. 
Together with Eq.(\ref{eqn:DROHF}), Eq.(\ref{eqn:SCFcond2c}) implies that $\lambda_{cv}=2\Delta_{cv}^{\rm UHF}$ at convergence. During the iterative procedure, we choose this same value for $\lambda_{cv}$ because it guarantees $\delta_s=0$ at each SCF cycle. Note that Eqs.(\ref{eqn:SCFcond2}) yield the SCF conditions for Roothaan's ROHF. Finally, our CUHF $\alpha$ and $\beta$ Fock matrices are 
\begin{eqnarray}
\tilde{\mathbf{F}}^\alpha = \left(
\begin{array}{ccc}
\mathbf{F}^\alpha_{cc}&\mathbf{F}^\alpha_{co}&\mathbf{F}^{\rm cs}_{cv}\\
\mathbf{F}^\alpha_{oc}&\mathbf{F}^\alpha_{oo}\:&\mathbf{F}^\alpha_{ov}\\
\mathbf{F}^{\rm cs}_{vc}&\mathbf{F}^\alpha_{vo}&\mathbf{F}^\alpha_{vv}
\end{array}
\right)
\:\:
\tilde{\mathbf{F}}^\beta = \left(
\begin{array}{ccc}
\mathbf{F}^\beta_{cc}&\mathbf{F}^\beta_{co}\:&\mathbf{F}^{\rm cs}_{cv}\\
\mathbf{F}^\beta_{oc}&\mathbf{F}^\beta_{oo}\:&\mathbf{F}^\beta_{ov}\\
\mathbf{F}^{\rm cs}_{vc}&\mathbf{F}^\beta_{vo}\:&\mathbf{F}^\beta_{vv}
\end{array}
\right).
\nonumber\\
\label{eqn:FROHFAB}
\end{eqnarray}
These CUHF Fock matrices are different from the UHF ones only in the $cv$ and $vc$ blocks, and are different from Roothaan's effective Fock matrix of Eq.(\ref{eqn:Roothaan}). Our CUHF procedure yielding ROHF is surprisingly straightforward: one simply performs UHF with Fock matrices replaced by Eqs.(\ref{eqn:FROHFAB}). 
These Fock matrices eliminate ambiguities arising in ROHF theory and produce a more physical UHF-like picture. 
In open-shell molecules, $\alpha$ and $\beta$ electrons feel different potentials; our $\tilde{\mathbf{F}}^\alpha$ and $\tilde{\mathbf{F}}^\beta$ operators are different from each other and yield $\alpha$ orbitals different from $\beta$ orbitals that are true ``canonical orbitals" obtained by diagonalization.
However, unlike UHF, they have no spin contamination, which is removed by Lagrangian constraints. Their eigenvalues $\varepsilon_i^\sigma$ are physical orbital energies in the sense that they are associated with individual $\alpha$ and $\beta$ orbitals, satisfy Koopmans' theorem, and the aufbau principle,\cite{Lieb} as opposed to many ROHF canonicalizations of Eq.(\ref{eqn:Roothaan}).\cite{GS10} Our orbitals have previously been proposed in the literature as semi-canonical orbitals for MP2 and used in an {\it ad hoc} fashion.\cite{RMP2} Our present work shows that the Fock matrices for which these orbitals are eigenfunctions appear from a constrained UHF optimization that eliminates spin contamination.

\begin{table}[t]
\caption{Mean and mean absolute errors of ionization potentials (-$\varepsilon_{\rm HOMO}$ in eV) of 24 open-shell systems. See the Supplementary Material for individual values.\cite{SuppM}
\label{tb:IPs}}
\begin{tabular}{crrr}
\hline\hline
 &   ROHF\footnote{Parameters from Ref. [\onlinecite{McWeeny}].}   &   UHF   &   CUHF    \\
\hline
ME          &    7.38 &   -0.68    &    -0.54  \\
MAE         &    7.38 &    0.71    &     0.61  \\
\hline\hline
\end{tabular}
\end{table}

{\it Results.} We have implemented CUHF in the \texttt{Gaussian} suite of programs\cite{GAUSSIAN} and verified that our procedure converges to the ROHF energy. Unlike many ROHF schemes, CUHF has no issues with SCF convergence. This is undoubtedly related to the observance of the aufbau principle in our method. Since Koopmans' theorem is valid for CUHF, orbital energies approximate ionization potentials (IP) and electron affinities (EA). In Table \ref{tb:IPs} we summarize the mean (ME) and mean absolute errors (MAE) of first IPs estimated via HOMO energies ($\varepsilon_{\rm HOMO}$) for 24 open-shell compounds selected from the G2 set.\cite{G2} Molecular geometries are optimized with B3LYP/6-31G(2df,p). CUHF results with a 6-311++G(3df,3pd) basis are compared to UHF and the default (McWeeny) ROHF implementation\cite{McWeeny} in \texttt{Gaussian}. In all systems, the CUHF $\varepsilon_{\rm HOMO}$ captures the right physics yielding IPs comparable to those of UHF yet preserving the correct $\langle S^2 \rangle$ expectation value.\cite{SuppM}
\begin{table}[t]
\caption{CN orbital energies in eV. In Roothaan's ROHF schemes (McWeeny and PGB), 5$\sigma$ is the open-shell orbital.
\label{tb:CN-orb}}
\begin{tabular}{lrrrrr}
\hline\hline
             &  3$\sigma$ & 4$\sigma$   &   1$\pi$  &  5$\sigma$ & 6$\sigma$  \\
\hline
McWeeny      &   -33.81   &   -16.89    &   -13.87  &   -6.21    &  1.94      \\
PGB          &   -33.46   &   -16.44    &   -13.68  &  -18.17    &  1.87      \\
CUHF $\alpha$&   -34.54   &   -19.87    &   -14.26  &  -16.06    &  1.87      \\
CUHF $\beta$ &   -33.46   &   -16.44    &   -13.68  &   -1.88    &  2.37      \\
UHF $\alpha$ &   -33.89   &   -20.20    &   -14.17  &  -15.47    &  1.87      \\
UHF $\beta$  &   -34.18   &   -16.91    &   -14.66  &   -1.06    &  2.42      \\
Exptl. IP\footnote{Ref.[\onlinecite{GS10}]}       &            &    15.5\;\;     &    14.4\;\;  &   14.2\;\;     &            \\
\hline\hline
\end{tabular}
\end{table}

\begin{table}[t]
\caption{ TDHF valence (V) and Rydberg (R) excitation energies (in eV) of open-shell molecules. Numbers in parentheses are UHF spin contamination $\delta_s$. \label{tb:TDHF}}
\begin{tabular}{llrrr}
\hline\hline
System	&	State	&	UHF	&	CUHF	&	Exptl.\\
\hline
BeF
			&    V $^2\Pi$ 	 &	4.20 &	4.19 &	4.14 \\
(0.001)	& R $^2\Sigma^+$	 &	6.34 &	6.33 &	6.16 \\
			& R $^2\Sigma^+$	 &	6.54 &	6.54 &	6.27 \\
BeH		&    V $^2\Pi$		 &	2.69 &	2.64 &	2.48 \\
(0.002)	&    R $^2\Pi$		 &	6.26 &	6.25 &	6.32 \\
CH3		&    R $^2$A$_1'$  &	6.54 &	6.23 &	5.73 \\
(0.012)	&    R $^2$A$_2''$ &	7.73 &	7.34 &	7.44 \\
CO$^+$& V $^2\Pi$			 &	6.93 &	4.84 &	3.26 \\
(0.141)	& V $^2\Sigma^+$	 &	11.10&	9.81 &	5.82 \\
CN			& V $^2\Pi$			 &	4.13 &	0.95 &	1.32 \\
(0.406)	& V $^2\Sigma^+$	 &	5.42 &	2.01 &	3.22 \\
\hline
ME			&						 &	1.43 &	0.45 &		  \\
MAE		&						 &	1.44 &	0.77 &		  \\
\hline\hline
\end{tabular}
\end{table}

We have compared our CUHF orbital energies with those obtained by  Eq.(\ref{eqn:Roothaan}) with parameters recently suggested by Plakhutin, Gorelik, and Breslavskaya (PGB).\cite{PGB} The PGB parametrization is chosen to obey Koopmans' theorem. However, the PGB scheme usually violates the aufbau principle resulting in poor SCF convergence. Therefore, as a simple remedy, we have used the converged ROHF wave function and then diagonalized Eq.(\ref{eqn:Roothaan}) with PGB parameters in a single shot. The eigenvalues thus obtained are identical to those from the self-consistent PGB scheme. 

For the systems in Table \ref{tb:IPs}, PGB gives the same first IP as CUHF, except for HCO whose $\varepsilon_{\rm HOMO}$ are $-10.88$ and $-10.40$ eV, respectively. The PGB scheme produces CUHF $\alpha$ virtual and $\beta$ core orbital energies by construction. However, the overall spectrum is appreciably different. In Table \ref{tb:CN-orb}, we present valence orbital energies for CN. In standard ROHF, 5$\sigma$ is predicted to be the open-shell orbital. Note the aufbau violation in PGB as previously reported.\cite{GS10} Results in Table \ref{tb:CN-orb} indicate that CUHF can well describe both $\alpha$ electron 
detachment and $\beta$ electron attachment processes, yielding a spectrum that is fully interpretable.

Last, we present excitation energies of five small open-shell molecules calculated with time-dependent HF (TDHF) based on UHF and CUHF with a 6-311++G(3df,3pd) basis. The bond-lengths for BeF and CO$^+$ (not included in the G2 set) are 1.355 and 1.078 $\mbox{\AA}$, respectively. For TD-CUHF, we have used CUHF orbitals and orbital energies in the TD-UHF procedure. Although this TD-CUHF scheme is not rigorous (one should perturb $\tilde{\mathbf{F}}$, compute the response of $\mathbf{P}$ and include terms arising from $\bm{\lambda}$), this simple approximation turns out to be quite reasonable as shown in Table \ref{tb:TDHF}. When UHF spin contamination ($\delta_s$) is small, TD-UHF and TD-CUHF give very similar results. As $\delta_s$ becomes larger, however, TD-UHF greatly overestimates the excitation energies. On the other hand, by retaining a spin projected reference ($\delta_s=0$), TD-CUHF gives more reasonable excitation energies outperforming TD-UHF in spin contaminated situations. 

This work was supported by NSF(CHE-0807194) and the Welch Foundation (C-0036). We thank Tom Henderson for a critical reading of this paper.

\newpage
{\bf Supplemental Material:}
\begin{table}[ht]
\caption{$\varepsilon_{\rm HOMO}$ of open-shell systems (in eV).
\label{tb:IPs-comp}}
\begin{tabular}{crrrr}
\hline\hline
Atom   &   ROHF\footnote{Parameters from Ref. [\onlinecite{McWeeny}].}   &   UHF   &   CUHF   &   Exptl. IP\footnote{Ref.[\onlinecite{G2}]}   \\
\hline
H           &   -3.40 &  -13.60    &   -13.60   &   13.60 \\
Li          &   -1.44 &   -5.34    &    -5.33   &    5.39 \\
B           &   -1.57 &   -8.67    &    -8.43   &    8.30 \\
C           &   -2.38 &  -11.95    &   -11.80   &   11.26 \\
N           &   -3.29 &  -15.55    &   -15.46   &   14.54 \\
O           &   -4.87 &  -14.21    &   -14.37   &   13.61 \\
F           &   -6.55 &  -18.54    &   -18.62   &   17.42 \\
Na          &   -1.35 &   -4.95    &    -4.95   &    5.14 \\
Al          &   -1.22 &   -5.94    &    -5.72   &    5.98 \\
Si          &   -1.98 &   -8.20    &    -8.09   &    8.15 \\
P           &   -2.85 &  -10.67    &   -10.66   &   10.49 \\
S           &   -4.07 &  -10.30    &   -10.11   &   10.36 \\
Cl          &   -5.40 &  -13.09    &   -13.00   &   12.97 \\
OH          &   -4.48 &  -13.98    &   -14.13   &   13.01 \\
PH$_2$      &   -2.89 &  -10.25    &    -9.94   &    9.82 \\
SH          &   -4.00 &  -10.35    &   -10.31   &   10.37 \\
NH          &   -3.25 &  -13.82    &   -13.79   &   13.49 \\
O$_2$       &   -3.86 &  -15.25    &   -14.52   &   12.07 \\
S$_2$       &   -3.34 &  -10.46    &   -10.05   &    9.36 \\
CH$_3$      &   -2.01 &  -10.46    &   -10.18   &    9.84 \\
C$_2$H$_5$  &   -1.65 &   -9.51    &    -9.25   &    8.12 \\
CN          &   -6.21 &  -14.17    &   -13.68   &   13.60 \\
HCO         &   -2.60 &  -10.73    &   -10.40   &    8.14 \\
CH3O        &   -3.93 &  -12.16    &   -12.29   &   10.73 \\
ME          &    7.38 &   -0.68    &    -0.54   &         \\
MAE         &    7.38 &    0.71    &     0.61   &         \\
\hline\hline
\end{tabular}
\end{table}
\end{document}